\magnification\magstep1
\parskip=\medskipamount
\hsize=6 truein
\vsize=8.2 truein
\hoffset=.2 truein
\voffset=0.4truein
\baselineskip=14pt

\font\titlefont=cmbx12
 at 10 truept
\font\authorfont=cmcsc10
\font\addressfont=cmsl10 at 10 truept
\font\smallbf=cmbx10 at 10 truept
 4
\font\tenssdb=cmssdc10

\def\Box#1{\mathop{\mkern0.5\thinmuskip
\vbox{\hrule
\hbox{\vrule
\hskip#1
\vrule height#1 width 0pt\vrule}%
\hrule}%
\mkern0.5\thinmuskip}}

\def\shalf{\hbox{${\textstyle{{1\over 2}}}$}}
\def\quarter{\hbox{${\textstyle{{1\over 4}}}$}}

\def\shalfalpha{\hbox{${\textstyle{{1\over 2\alpha}}}$}}
\def\overalpha{\hbox{${\textstyle{{1\over\alpha}}}$}}
\def\reals{{\Bbb R}}

\def\equaltwo{\mathrel{\hbox{${\displaystyle{\mathop{=}^2}}$}}}

\def\G{\hbox{{\tenssdb G}}}
\def\LG{\hbox{{\tenssdb LG}}}
\def\gfour{\hbox{$g^{(4)}$}}
\def\Rfour{\hbox{$R^{(4)}$}}
\def\M{{\cal M}}
\def\P{{\cal P}}
\def\hamc{{\cal H}}
\def\diffc{{\cal D}}
\def\gaussc{{\cal G}}
\def\first{1^{\underline{\rm st}}}
\def\second{2^{\underline{\rm nd}}}
\def\muth{\mu^{\underline{\rm th}}}
\def\t#1{\tilde{#1}}
\def\beginproof{\noindent{\authorfont Proof.\hskip 0.35em \relax}}
\def\endproof{\Box{5pt}}
\def\Bbb#1{\bf #1}                                                    

\outer\def\beginsection#1\par{\vskip0pt plus.2\vsize\penalty-150
\vskip0pt plus-.2\vsize\vskip1.2truecm\vskip\parskip
\message{#1}\leftline{\bf#1}\nobreak\smallskip\noindent}

\newcount\notenumber

\def\note{\advance\notenumber by 1
\footnote{$^{\{\the \notenumber\}}$}}

\newdimen\itemindent \itemindent=13pt
\def\textindent#1{\parindent=\itemindent\let\par=\resetpar%
\indent\llap{#1\enspace}\ignorespaces}

\let\oldpar=\par
\def\resetpar{\oldpar\parindent=0pt\let\par=\oldpar}

\font\ninerm=cmr9 \font\ninesy=cmsy9
\font\eightrm=cmr8 \font\sixrm=cmr6
\font\eighti=cmmi8 \font\sixi=cmmi6
\font\eightsy=cmsy8 \font\sixsy=cmsy6
\font\eightbf=cmbx8 \font\sixbf=cmbx6
\font\eightit=cmti8
\def\eightpoint{\def\rm{\fam0\eightrm}
  \textfont0=\eightrm \scriptfont0=\sixrm \scriptscriptfont0=\fiverm
  \textfont1=\eighti  \scriptfont1=\sixi  \scriptscriptfont1=\fivei
  \textfont2=\eightsy \scriptfont2=\sixsy \scriptscriptfont2=\fivesy
  \textfont3=\tenex   \scriptfont3=\tenex \scriptscriptfont3=\tenex
  \textfont\itfam=\eightit  \def\it{\fam\itfam\eightit}%
  \textfont\bffam=\eightbf  \scriptfont\bffam=\sixbf
  \scriptscriptfont\bffam=\fivebf  \def\bf{\fam\bffam\eightbf}%
  \normalbaselineskip=9pt
  \setbox\strutbox=\hbox{\vrule height7pt depth2pt width0pt}%
  \let\big=\eightbig \normalbaselines\rm}
\catcode`@=11 %
\def\eightbig#1{{\hbox{$\textfont0=\ninerm\textfont2=\ninesy
  \left#1\vbox to6.5pt{}\right.\n@space$}}}
\def\vfootnote#1{\insert\footins\bgroup\eightpoint
  \interlinepenalty=\interfootnotelinepenalty
  \splittopskip=\ht\strutbox %
  \splitmaxdepth=\dp\strutbox %
  \leftskip=0pt \rightskip=0pt \spaceskip=0pt \xspaceskip=0pt
  \textindent{#1}\footstrut\futurelet\next\fo@t}
\catcode`@=12 %


\rightline{ZU-TH 9/98}
\rightline{gr-qc 9805065}
\bigskip
{\baselineskip=24 truept
\titlefont
\centerline{THE GENERALIZED THIN-SANDWICH PROBLEM}
\centerline{AND ITS LOCAL SOLVABILITY}
}

\vskip 1.1 truecm plus .3 truecm minus .2 truecm

\centerline{\authorfont Domenico Giulini}
\vskip 2 truemm
{\baselineskip=12truept
\addressfont
\centerline{Institut f\"ur theoretische Physik, Universit\"at Z\"urich}
\centerline{Winterthurerstrasse 190, CH-8057 Z\"urich, Switzerland}
\centerline{e-mail: giulini@physik.unizh.ch}
}
\vskip 1.5 truecm plus .3 truecm minus .2 truecm

\centerline{\smallbf Abstract}
\vskip 1 truemm
{\baselineskip=12truept
\leftskip=3truepc
\rightskip=3truepc
\parindent=0pt

{\eightpoint
We consider Einstein Gravity coupled to matter consisting of a 
gauge field with any compact gauge group and coupled scalar fields.
We investigate under what conditions a free specification of a 
spatial field configuration and its time derivative determines a 
solution to the field equations (thin-sandwich problem). We establish 
sufficient conditions under which the thin-sandwich problem can 
be solved locally in field space.
\par}}

\beginsection{Introduction}

In this paper we consider the initial value problem for 
Einstein gravity plus matter in spacetimes $\Sigma\times\reals$, 
where $\Sigma$ is a closed orientable 3-manifold. We are 
interested in the question of how to find initial data which
satisfy the constraints. The most popular approach here is 
a powerful method  devised by Lichnerowicz, Choquet-Bruhat, York and 
others, henceforth referred to as the ``conformal method''.
(See [I] for a brief review and [CY] for more details.)
Of the gravitational variables it allows to freely specify the 
conformal class of the initial 3-metric, the conformally rescaled 
transverse-traceless components of the extrinsic curvature and a 
constant trace thereof (i.e. $\Sigma$ must have constant mean 
curvature).  Given these data, the constraints turn into a 
quasilinear elliptic system of second order for the conformal 
factor (scalar function) and the transverse momentum (vector field),
which decouples due to the constant mean-curvature condition. 
The disadvantages of this method are that it does not easily 
generalize to data of variable mean curvature and that it does 
not allow to control the local scales of the physical quantities 
initially, since the freely specifiable data (gravitational and 
non-gravitational) are related to the actual physical quantities 
by some rescalings with suitable powers of the conformal factor. 
In particular, one has no control over the conformal part of the 
initial 3-geometry. 

In this paper we are concerned with the so called ``thin-sandwich 
method'', which differs from the one just mentioned insofar as it 
aims to define solutions to the Einstein equations by a {\it free} 
specification of the initial field configuration and its 
coordinate-time-derivative. The constraints are now read as equations 
for the gauge parameters (lapse, shift, ..). From the conformal point 
of view this means that one tries to trade in the freedom to specify 
the gauge parameters for the freedom to specify the conformal part 
of the metric and the longitudinal part of the momentum.
The disadvantages mentioned above would then be overcome, but 
unfortunately the equations (for the gauge parameters) turn 
out to be non-elliptic\note{By ellipticity of non-linear 
differential operators one means the ellipticity of its 
linearization, which depends on the point (in field space) about 
which one linearizes. The usual statement that the thin-sandwich 
equations are not elliptic merely asserts the existence of points 
where the linearization is not elliptic, but not that the domain of 
ellipticity is empty.} in general~[BO][CF]. However, for certain 
open subsets of initial data they are elliptic and can be locally 
solved. This was first shown in [BF] and will be shown in a wider 
context with dynamical matter here.

We note that historically this approach arose from the question 
(then formulated as a conjecture) of whether the specification of two 
3-geometries uniquely determine an interpolating Einstein space-time 
(thick-sandwich problem). For nearby geometries infinitesimally 
close in time this turns into the thin-sandwich problem (see [W] 
(chapter~4) and [BSW]) which we now describe in more detail.

In a space-time neighborhood $\Sigma\times\reals$ of the Cauchy 
surface $\Sigma$, we use the standard parametrization of the 
space-time metric $\gfour$,
$$
\gfour=-\alpha^2\, dt\otimes dt + g_{ab}(dx^a+\beta^a dt)\otimes
                                   (dx^b+\beta^b dt),
\eqno{(1.1)}
$$
where $g$ is the ($t$-dependent) Riemannian metric of $\Sigma$ and 
$\alpha,\beta$ are ($t$-dependent) scalar- and vector fields 
on $\Sigma$, known as lapse and shift. The extrinsic curvature
reads
$$
K=\shalfalpha(\partial_{t}-L_{\beta})g,
\eqno{(1.2)}
$$
where $L_{\beta}$ denotes the Lie-derivative along $\beta$.
Written in terms of $(g,K)$, the constraints do not 
depend on $\alpha$ and $\beta$ and hence constrain the set 
of allowed values for the data $(g,K)$. The constraints read
$$\eqalignno{
& K_{ab}K^{ab}-(K_a^a)^2-R=-2T_{\perp\perp},   &(1.3)\cr
& \nabla^b(K_b^a-\delta_b^a K_c^c)=T_{\perp}^a,&(1.4)\cr}
$$
where $T$ is the energy-momentum tensor of the matter and 
$\perp$ denotes the component along the future pointing 
normal $n$ of $\Sigma$.

Alternatively, one may write the constraints in terms of $g$ and 
$\dot g:=\partial_t g$ by replacing $K$ via (1.2). Then they 
explicitly involve $\alpha$ and $\beta$ and one may ask whether 
it is possible to {\it freely} specify $(g,\dot g)$ and let
the constraints determine $\alpha$ and $\beta$.
This is the thin-sandwich problem (TSP). If we abbreviate
$\Psi:=(g,\dot g)$, $X:=(\alpha,\beta)$, the constraints take 
the form of the thin-sandwich equation (TSE): 
$$
F[\Psi,X]=0.
\eqno{(1.5)}
$$
The TSP now asks for existence and uniqueness of solutions of the TSE,
read as equation for $X$ given $\Psi$. Once this is solved, we can 
construct $(g,K)$ satisfying (1.3-4), which uniquely determine 
space-time via the Einstein evolution equations~[CY].

It has long been discussed in the literature that, in general, 
existence and uniqueness should fail~[BO], although some 
arguments merely showed this under the additional assumption 
(a priori) of constant lapse function $\alpha$~[CF] (see also [G] for a 
related issue). However, more recently 
it was shown that given a solution $(\Psi,X)$ of the TSE which 
satisfies certain bounds on geometric quantities and which admits 
no nontrivial solutions of the spatially projected Killing equation, 
there exist unique solutions $X'(\Psi')$ for all $\Psi'$ in a 
neighbourhood of $\Psi$. This was achieved by an 
implicit-function-theorem for a reduced version of (1.5) with 
already eliminated lapse function. 

Note that in this formulation of the TSP the right hand sides of the 
constraints, that is $T_{\perp\perp}$ and $T_{\perp}^a$, are assumed 
given. These are {\it not} the components of $T$ that an observer 
along $\partial_t$ would measure. The relation between the two 
sets of components involve $\alpha$ and $\beta$. For example, 
if the matter is represented by some dynamical field $\phi$, 
the quantities $T_{\perp\perp}$ and $T_{\perp}^a$ cannot be 
calculated from the initial data $(\phi,\dot \phi)$ without the 
use of $\alpha$ and $\beta$. Hence there is a certain inconsistency 
in the traditional formulation of the TSP, in that it eliminates 
any appearance of the normal $n$ in favour of $\partial_t$, lapse 
and shift on the left side (the gravitational part), but not on 
the right side (matter part) of the constraints. This inconsistency 
was already felt by others (see e.g. section IV of~[BO]), but no 
alternative formulation was hitherto attempted.  

In this paper we consider a generalized thin-sandwich problem (GTSP) 
for the initial data of the full system of coupled gravitational and
matter fields, which avoids the difficulty just mentioned.
As matter we shall consider coupled systems of scalar and gauge fields, 
further specified below. By $\Phi$ we shall collectively denote all 
dynamical fields of the theory. We ask: Under what conditions do input 
data $\Psi:=(\Phi,\dot\Phi)$ uniquely specify a solution to the 
Einstein-matter equations? In the same fashion as for (1.5), 
one obtains a now {\it generalized} thin-sandwich equation (GTSE) 
from which one tries to determine the ``gauge parameters'' $X$
given the data $\Psi$. One non-trivial aspect of our 
generalization is due to the possible presence of gauge matter fields. 
In this case $X$ comprises lapse, shift and {\it additional} 
functions with values in the Lie-algebra of the gauge group. Our main 
result will consist of an implicit-function-theorem for this extended 
set of variables, which for our GTSP is precisely analogous 
to the result proven in~[BF] for the traditional form of the 
thin-sandwich problem. But note that the two formulations differ even 
without gauge fields.

\beginsection{The Generalized Framework}

The dynamical fields we consider involve the gravitational field,
a gauge field with compact gauge group $\G$ of dimension $N$ and an 
$M$-component scalar field with values in an associated 
$\reals^M$-vector-bundle. It couples to both previous fields in the
standard minimal fashion. For simplicity we assume the $\G$-principal
bundle to be trivial. Since the frame bundle of any orientable 
3-manifold is always trivial, we may once and for all choose global 
trivialisations of these bundles and represent fields by their 
globally defined component-fields on space-time. For fixed time, 
a configuration of fields is given by the $6+3N+M$ component-fields 
on $\Sigma$,
$$
\Phi^A:=(g_{ab},A_a^{\mu},\phi^{\alpha}),
\eqno{(2.1)}
$$
where indices $\mu,\nu ..$ denote components in the Lie-algebra 
$\LG$ and $\alpha,\beta ..$ denote components in $\reals^M$.
Hence we think of a field configuration as mapping 
$$
\Phi:\Sigma\longrightarrow GL(3,\reals)/SO(3)\times
     \reals^{3N}\times\reals^M ,
\eqno{(2.2)}
$$
where we identify the space of symmetric, positive definite matrices,  
in which $g_{ab}$ is valued, with the first factor space on the right 
hand side. The total target space, whose dimension is $6+3N+M$, 
will be denoted by $\Theta$, and the space of mappings 
$\Sigma\rightarrow\Theta$ (to be further specified) by $\M$.


Compactness of the gauge-group $\G$ implies the existence of 
$\G$-invariant, symmetric, positive definite, bilinear forms 
$k_{\mu\nu}$ and $h_{\alpha\beta}$ on $\LG$ and $\reals^M$ respectively.
The class of models we shall consider here are characterized by the 
Lagrange four-form 
$$
{\cal L}=\shalf *\Rfour-\quarter k_{\mu\nu}\Omega^{\mu}\wedge *
\Omega^{\nu}-\shalf h^{\alpha\beta}\nabla\phi_{\alpha}\wedge *\nabla
          \phi_{\beta}-W,
\eqno{(2.3)}          
$$
where $*$ is the Hodge-duality map wrt. $g^{(4)}$ and $\Omega$ is the 
curvature of $A$. For notational simplicity we shall denote all the 
covariant derivatives
acting on sections in the various vector bundles by the same symbol 
$\nabla$. In general it will therefore involve the Christoffel symbols 
of $g$ as well as the gauge connection $A$ in the 
appropriate representation of $\LG$. The potential $W$ depends on the 
fields and its first spatial derivatives. Its precise form is not 
important, except that we need to explicitly exclude $\second$ (or higher) 
derivative couplings, in particular, the so called conformal coupling 
of the scalar and gravitational field. The reason for this will be
explained below. The Hamiltonian constraint for (2.3) has the general 
form
$$
\hamc=\shalf G_{AB}V^AV^B+U=0,
\eqno{(2.4)}
$$
where the potential is the following sum: 
$$
U=-R+\quarter k_{\mu\nu}g^{ac}g^{bd}\Omega_{ab}^{\mu}\Omega_{cd}^{\nu}
    +h_{\alpha\beta}g^{ab}\nabla_a\phi^{\alpha}\nabla_b\phi^{\beta}+W,
\eqno{(2.5)}
$$
whose terms represent the contributions of the gravitational, gauge, and 
scalar fields (two terms) respectively to the potential energy.
$R$ denotes the Ricci-scalar for $g$. The ``canonical velocities'', 
$V^A$, can be written in terms of the ``coordinate velocities'' 
${\dot\Phi}^A:=\partial_t\Phi^A$ and the ``gauge-parameters'' 
$\alpha$ and $\xi:=(\beta,\lambda)$. The general structure is 
(compare (1.2))
$$
V=\overalpha\Gamma
   =\overalpha\left(\dot\Phi+f_{\xi}\right),
\eqno{(2.6)}
$$
where $f_{\xi}$ represents the motion generated by the infinitesimal 
diffeomorphism and gauge-transformation with parameters $\beta$ and 
$\lambda$ respectively. Resolved in terms of the individual fields 
(2.1), the components of $\Gamma$, and hence of $f_{\xi}$, read:
$$\eqalignno{
\Gamma_{ab}&={\dot g}_{ab}-2\nabla_{(a}\beta_{b)}\,,
&(2.7)\cr
\Gamma_a^{\mu}&={\dot A}_a^{\mu}-\beta^b\Omega^{\mu}_{ba}
                -\nabla_a\lambda^{\mu}\,,
&(2.8)\cr
\Gamma^{\alpha}&=
{\dot \phi}^{\alpha}-\beta^a\nabla_a\phi^{\alpha}
+\lambda^{\mu}\rho^{\alpha}_{\mu\beta}\phi^{\beta},
&(2.9)\cr}
$$
where $\rho$ denotes the representation of $\LG$ in $gl(M,\reals)$. 

Finally, following the decomposition of $\Theta$ as cartesian 
product, the ``kinetic-energy-metric'' $G_{AB}$ on $\Theta$ 
which appears in (2.4) has the following block-structure:
$$
G_{AB}=G^{ab\, cd}\oplus k_{\mu\nu}g^{ab}\oplus h_{\alpha\beta}, 
\eqno{(2.10)}
$$
where the first $6\times 6$ block is given by the DeWitt-metric
$$
G^{ab\, cd}=\quarter\left(g^{ac}g^{bd}+g^{ad}g^{bc}-2g^{ab}g^{cd}\right),
\eqno{(2.11)}
$$
which is a Lorentz metric of signature (1,5). Hence $G_{AB}$ itself 
is a Lorentz metric of signature $(1,5+3N+M)$ on the manifold
$\Theta$, which is homeomorphic to $\reals^{6+3N+M}$. We shall 
sometimes denote this metric simply by $G$ and write $G(\cdot,\cdot)$ 
for the inner product. We will see that the Lorentzian signature 
of $G$ is {\it the} important feature on which the proofs of our 
main results rest. This is also the reason why we had to exclude 
higher derivative (e.g.~conformal) couplings of the scalar and 
gravitational fields, since they will in general destroy this 
signature structure~[K]. On the other hand, our proofs will 
still apply to more complicated self-couplings of the scalar field. 
 For example, non-linear $\sigma$-models would be allowed,
since here the target space metric of the scalar field, 
$h_{\alpha\beta}$, simply becomes $\phi^{\alpha}$-dependent, 
which is unimportant to our proofs as long as it stays positive 
definite.

The (undensitized) momenta of the field $\Phi^A$ are just given by 
the covariant components -- with respect to $G$ -- of the 
velocities: $P_A:=G_{AB}V^B$. For the individual fields we write 
$P_A=(\pi^{ab},\pi_{\mu}^a, \pi_{\alpha})$. In the canonical theory,
the phase space function that generates infinitesimal diffeomorphisms 
and gauge-transformations with parameter $\xi'$ is given by 
$$
\P_{\xi'}:=\int_{\Sigma}d\mu\,P_Af^A_{\xi'},
\eqno{(2.12)}
$$
where here and below we set $d\mu:=\sqrt{\hbox{det}\{g_{ab}\}}d^3x$ 
for the measure on $\Sigma$. For completeness we remark that, using 
(2.7-9), a straightforward calculation yields the following 
familiar Poisson-bracket relation, which involves the curvature 
tensor $\Omega$ of the gauge field on $\Sigma$: 
$$
\left\{\P_{\xi'},\P_{\xi''}\right\}=P_{\xi'''},
\eqno{(2.13)}
$$
where $\xi'''=(\beta''',\lambda''')$ reads
$$\eqalignno{
\beta'''&=-[\beta',\beta''],
&(2.14)\cr
\lambda'''&=[\lambda',\lambda'']-\Omega(\beta',\beta'').
&(2.15)\cr}
$$

The diffeomorphism- and Gau\ss -constraints are just given by
$\P_{\xi'}=0\,\forall \xi'$, which may be expressed by saying
that the velocity field $V$ is $L^2G$-orthogonal to all ``vertical'' 
vector fields $f_{\xi'}$. Hence we must have
$$
0  =  \int_{\Sigma}P_Af^A_{\xi'}\, d\mu 
   =  \int_{\Sigma}\overalpha G(\Gamma,f_{\xi'})\, d\mu 
   =: \int_{\Sigma}(g_{ab}{\beta'}^a\diffc^b+
                  k_{\mu\nu}{\lambda'}^{\mu}\gaussc^{\nu})\,d\mu,
\eqno{(2.16)}
$$
for all, say $C^{\infty}$, vector fields $\beta'$ and $\LG$-valued 
functions $\lambda'$. This is equivalent to 
$\diffc^a=0$ (diffeomorphism constraint) and 
$\gaussc^{\mu}=0$ (Gau\ss\ constraint). Explicitly we get
$$\eqalignno{
&\diffc^a=2G^{abcd}\nabla_b\left(\overalpha\Gamma_{cd}\right)
-\overalpha g^{ab}k_{\mu\nu}\Omega^{\mu}_{bc}g^{cd}\Gamma_d^{\nu}
-\overalpha g^{ab}(\nabla_b\phi^{\alpha})h_{\alpha\beta}\Gamma^{\beta}\,,
&(2.17)\cr
&\gaussc^{\mu}=\nabla_a\left(\overalpha g^{ab}\Gamma^{\mu}_b\right)
+\overalpha k^{\mu\nu}\rho^{\alpha}_{\nu\beta}\phi^{\beta}
h_{\alpha\gamma}\Gamma^{\gamma}\,.
&(2.18)\cr}
$$

Given $\Phi$ and $\dot\Phi$ we now have the $4+N$ equations 
$0=\hamc=\diffc^a=\gaussc^{\mu}$ for the $4+N$ unknowns 
$\alpha,\beta^a,\lambda^{\mu}$. The first step consists of 
inserting (2.6) into (2.4) and solving for $\alpha^2$:
$$
\alpha^2=-{G(\Gamma,\Gamma)\over 2U}.
\eqno{(2.19)}
$$
 For this to make sense the right-hand side must be positive. 
But for the following analysis it turns out that we need to 
put the following stronger 

\noindent
{\bf Condition 1 (a priori):}
$$\eqalignno{
U&>0,  &(2.20)\cr
G(\Gamma,\Gamma)&<0. &(2.21)\cr}
$$
Note that (2.20) just involves the initial data, whereas (2.21) 
contains as well the unknowns $\beta$ and $\lambda$ (hence ``a priori''). 
Note that (2.21) says that the system must, at each point of $\Sigma$, 
move in a ``timelike'' direction with respect to the Lorentz metric 
$G$. The need for such an a priori bound implies that our results 
will only be perturbative. Given these bounds, we set $\alpha$ equal to 
the positive square root of the right hand side of (2.19). 
We can then eliminate $\alpha$ from (2.17),(2.18) and obtain 
a set of $3+N$ equations for the $3+N$ unknowns $\xi=(\beta,\lambda)$, 
which we call the generalized {\it reduced} thin-sandwich 
equation (GRTSE):
$$
F[\Psi,\xi]=0.
\eqno{(2.22)}
$$

Now consider the following functional over configurations satisfying 
Condition~1:
$$
S[\Psi,\xi]=\int_{\Sigma}\sqrt{-2U\,G(\Gamma,\Gamma)}\,d\mu,
\eqno{(2.23)}
$$
then solutions to the GRTSE are stationary points with respect to 
variations in $\xi$. More precisely, let $D_2$ denote the partial 
(functional-) derivative in the second argument and denote the 
$L^2$ inner product of $\xi_1$ and $\xi_2$ by 
$\langle\xi_1\vert\xi_2\rangle:=\int_{\Sigma}d\mu
(k_{\mu\nu}{\lambda_1}^{\mu}{\lambda_2}^{\nu}+g_{ab}\beta_1^a\beta_2^b)$.
We then have

\proclaim Lemma 1.
$$
D_2S[\Psi,\xi](\xi')=-\langle\xi'\vert F[\Psi,\xi]\rangle
\eqno{(2.24)}
$$

\beginproof
 For $s\in (-\epsilon,\epsilon)$ set $\eta(s)=\xi+s\xi'$ and 
$\Gamma(s)=\dot\Phi+f_{\eta(s)}$. Hence 
${d\over ds}\vert_{s=0}\Gamma(s)=f_{\xi'}$. 
Then, recalling (2.16), 
$$
{d\over ds}\Big\vert_{s=0} S[\Psi,\eta(s)]
=-\int_{\Sigma}\sqrt{-2U\over G(\Gamma,\Gamma)}\ G(\Gamma,f_{\xi'})\, d\mu
=-\langle\xi'\vert F[\Psi,\xi]\rangle
\eqno{(2.25)}
$$
\vskip -1.0truecm
\hfill
$\endproof$

\beginsection{Main Results}

In this section we are mainly concerned with the linearization of the 
GRTSE, except at the end where we will discuss global uniqueness. 
The corresponding linear operator will be called $L$, without explicit 
indication that it depends on $\Psi$ and $\xi$. It is defined by
$$
\langle\xi'\vert L\xi''\rangle
:={d\over ds}\Big\vert_{s=0}\langle\xi'\vert F[\Psi,\xi+s\xi'']\rangle
 = \int_{\Sigma}d\mu\ {d\over ds}\Big\vert_{s=0}
    {G(f_{\xi'},\Gamma(s))\over\alpha(s)} ,
\eqno{(3.1)}
$$
where $\Gamma(s)=\dot\Phi+f_{\xi+s\xi''}$ and 
$\alpha(s)=[-G(\Gamma(s),\Gamma(s))/2U]^{1/2}$. Setting 
$\Gamma(s=0)=\Gamma$, $\alpha(s=0)=\alpha$, and noting that 
${d\over dt}\big\vert_{s=0}\Gamma(s)=f_{\xi''}$, we get
$$
{d\over ds}\Big\vert_{s=0}\alpha(s)=-{G(f_{\xi''},\Gamma)\over 2\alpha U}
=\alpha\,{G(f_{\xi''},\Gamma)\over G(\Gamma,\Gamma)},
\eqno{(3.2)}
$$
where we used (2.4) to eliminate $U$ in the last step. Hence
$$
{d\over ds}\Big\vert_{s=0}{G(f_{\xi'},\Gamma(s))\over\alpha(s)}
={1\over\alpha}\left[G(f_{\xi'},f_{\xi''})
 -{G(f_{\xi'},\Gamma)G(f_{\xi''},\Gamma)\over G(\Gamma,\Gamma)}\right]
={G(f^{\perp}_{\xi'},f^{\perp}_{\xi''})\over\alpha},
\eqno{(3.3)}
$$
where
$$
f^{\perp}_{\xi'}:=f_{\xi'}-\Gamma{G(f_{\xi'},\Gamma)\over G(\Gamma,\Gamma)}
\eqno{(3.4)}
$$
is the $G$-orthogonal projection of $f_{\xi'}$ perpendicular to 
$\Gamma$. This leads to the following expression for $L$'s matrix
elements:
$$
\langle\xi'\vert L\xi''\rangle=\int_{\Sigma}d\mu\,
\overalpha G(f^{\perp}_{\xi'},f^{\perp}_{\xi''})\,, 
\eqno{(3.5)}
$$
where $\alpha$ is the square-root of the rhs. of (2.19).
If (2.21) holds, $\Gamma$ is ``timelike'' (pointwise on $\Sigma$)
and hence the $f^{\perp}$'s are ``spacelike'' or zero. Since the metric 
$G$ is Lorentzian, it is positive definite on ``spacelike'' vectors. 
Hence we have shown

\proclaim Lemma 2. Suppose Condition~1 holds, then $L$ 
is self-adjoint and non-negative. Furthermore,
$\xi'\in\,\hbox{kernel}\, (L)\Leftrightarrow\exists\,\kappa:
\Sigma\rightarrow\reals$
such that
$$
f_{\xi'}=\kappa\Gamma .
\eqno{(3.6)}
$$

\noindent
{\it Remark:} Symmetry is expected, for consider the rhs. of 
(2.23) as functional of $\Psi$ and $\Gamma$, denoting it by 
$S[\Psi,\Gamma]$, then the calculation of the rhs. of (3.5) was 
just that of $-D_2^2S[\Psi,\Gamma](\Gamma',\Gamma'')$ for 
$\Gamma'=f_{\xi'}$ and $\Gamma''=f_{\xi''}$. 

We will next show that under the same hypotheses $L$ is in fact
elliptic. For this we will need

\proclaim Lemma 3. If Condition~1 holds, $\pi^{ab}$ 
is a positive or negative definite matrix.

\beginproof
Condition 1 implies
$G^{AB}P_AP_B<0\Rightarrow G_{ab\, cd}\pi^{ab}\pi^{cd}=
2(\pi_{ab}\pi^{ab}-\shalf(\pi^a_a)^2)<0$. Choosing a frame where 
$g_{ab}=\delta_{ab}$ and $\pi^{ab}=\hbox{diag}\,(p_1,p_2,p_3)$,
this is equivalent to the following condition on the eigenvalue-vector 
$\vec p:=(p_1,p_2,p_3)$: 
$$
\Big\vert{\vec n\cdot\vec p\over\Vert\vec p\Vert}\Big\vert 
>\sqrt{2\over 3}\,,
\eqno{(3.7)}
$$
where $\vec n=(1,1,1)/\sqrt{3}$, which means that $\vec p$ 
lies in the interior of the double cone with axis along $\vec n$
and opening angle $\theta<{\cos}^{-1}(2/3)^{1/2}$ about its axis.
This cone just touches the walls of the positive and negative 
octants along the bisecting lines, and since $\vec p$ must be in 
its interior, all eigenvalues are either strictly positive or 
strictly negative. \hfill 
$\endproof$

\proclaim Proposition 1. Suppose Condition~1 holds, then 
the second order differential operator $L$ is elliptic.

\beginproof
We shall calculate the principal symbol of $L$.
 For this we need to go back to the explicit formulae (2.17) 
and (2.18) for the full non-linear problem and explicitly 
linearize them, but keeping track only of the highest (second) 
derivatives. By $\equaltwo$ we shall denote equality 
in the second derivative terms. We set again 
$\Gamma(s)=\dot\Phi+f_{\xi+s\xi'}$ etc., with $\xi'=(\beta',\lambda')$.
It will be convenient to express things in terms of the momenta, using
$\alpha P_A =G_{AB}\Gamma^B$, and accordingly write (3.2) in the form
$$
{d\over ds}\Big\vert_{s=0}\alpha(s)=-{P_Af^A_{\xi'}\over 2U}.
\eqno{(3.8)}
$$
Then
$$\eqalignno{
&{d\over ds}\Big\vert_{s=0}\diffc^a
\equaltwo\partial_b \left[{P_Af^A_{\xi'}\over\alpha U}\pi^{ab}
-{4\over\alpha}G^{ab\, cd}\partial_c\beta'_d\right]
&\cr
&\quad\equaltwo -{1\over\alpha}\left[{2\pi^{ac}\pi^{de}g_{db}
\partial_c\partial_e{\beta'}^b+
\pi^{ab}\pi^c_{\nu}\partial_b\partial_c{\lambda'}^{\nu}\over U}
+g^{bc}\partial_b\partial_c{\beta'}^a
-g^{ac}\partial_c\partial_b{\beta'}^b\right]
&(3.9)\cr
&&\cr
&{d\over ds}\Big\vert_{s=0}\gaussc^{\mu}
\equaltwo\partial_a\left[{P_Af^A_{\xi'}\over 
2\alpha U}k^{\mu\nu}\pi^a_{\nu}-
{1\over\alpha}g^{ab}\partial_b{\lambda'}^{\mu}\right]
&\cr
&\quad\equaltwo -{1\over\alpha}\left[{2k^{\mu\nu}\pi_{\nu}^ag_{bc}\pi^{cd}
\partial_a\partial_d{\beta'}^b
+k^{\mu\sigma}\pi_{\sigma}^a\pi_{\nu}^b
\partial_a\partial_b{\lambda'}^{\nu}\over 2U}
+g^{ab}\partial_a\partial_b{\lambda'}^{\mu}\right].
&(3.10)\cr}
$$
Replacing $\partial_a\rightarrow k_a$ we can just read-off the  
matrix of the principal symbol $\sigma(k)$ in the general form 
$$
\def\linie{\vrule height 8pt depth 5pt width 0.4pt}
\lineskip=0pt
\left[\matrix{\sigma^{\mu}_{\,\nu} & \linie & \sigma^{\mu}_{\,b}\cr
\noalign{\hrule}              
              \sigma^a_{\,\nu}     & \linie & \sigma^a_{\,b}    \cr}
\right],
\eqno{(3.11)}
$$
where we have chosen to order the $N+3$ rows and columns so that we 
first count the $N$ components of $\lambda'$ and then the 3~components 
of $\beta'$. In order to calculate the determinant we make the following 
simplifications: We call $\pi^{ab}k_b=:p^a$, $\pi_{\mu}^ak_a=:\pi_{\mu}$,
$\pi^{\mu}:=k^{\mu\nu}\pi_{\nu}$
and choose a spatial frame where $g_{ab}=\delta_{ab}$, 
$k_a=(\Vert k\Vert,0,0)$ 
and $p^a=(p_1,p_2,0)$. Then (3.11) reads explicitly
$$
\def\linie{\vrule height 12pt depth 5pt width 0.4pt}
\def\back{\noalign{\vskip-2pt}}
\lineskip=0pt
\sigma(k)=-{1\over\alpha}
\left[\matrix{{\pi^{\mu}\pi_{\nu}\over 2U}+
\Vert k\Vert^2\delta^{\mu}_{\nu}
&\linie & {\pi^{\mu}p_1\over U} & {\pi^{\mu}p_2\over U} & 0
\cr
\noalign{\hrule}
{p_1\pi_{\nu}\over U}
&\linie & {2p_1^2\over U} & {2p_1p_2\over U} & 0
\cr\back
{p_2\pi_{\nu}\over U}
&\linie & {2p_1p_2\over U} & {2p_2^2\over U}+\Vert k\Vert^2 & 0
\cr\back
0&\linie & 0&0&\Vert k\Vert^2\cr}
\right].
\eqno{(3.12)}
$$
Now, for $k\not =0$, $p_1=\Vert k\Vert\pi(\hat k,\hat k)$, where
$\hat k:=k/\Vert k\Vert$. Lemma~3 then ensures that $p_1\not =0$.
In order to calculate $\hbox{det}\{\sigma(k)\}$, we simplify this matrix
as follows: We subtract ${\pi^{\mu}\over 2p_1}$ times the $N+\first$ 
row from the $\muth$ row, for each $1\leq\mu\leq N$, and also subtract 
${p_2\over p_1}$ times the $N+\first$ row from the $N+\second$ row. The 
resulting matrix reads 
$$
\def\linie{\vrule height 12pt depth 5pt width 0.4pt}
\def\back{\noalign{\vskip -2pt}}
\lineskip=0pt
-{1\over\alpha}\left[\matrix{
\Vert k\Vert^2\,\delta_{\nu}^{\mu} & \linie & 0 & 0 & 0 \cr
\noalign{\hrule}
{p_1\pi_{\nu}\over U} & \linie & {2p_1^2\over U} & 
{2p_1p_2\over U} & 0 \cr\back
0 & \linie & 0 & \Vert k\Vert^2 & 0 \cr\back
0 & \linie & 0 & 0 & \Vert k\Vert^2 \cr}\right].
\eqno{(3.13)}
$$
Its determinant, which equals that of $\sigma(k)$, is now easily 
calculated:
$$
\hbox{det}\{\sigma(k)\}=
{2p_1^2\over U}\Vert k\Vert^{2(N+2)}\left[-{1\over\alpha}\right]^{N+3}
=2\left[-{\Vert k\Vert^2\over\alpha}\right]^{N+3}
{[\pi(\hat k,\hat k)]^2\over U}.
\eqno{(3.14)}
$$
Lemma~3 implies that this is zero $\Leftrightarrow$ $k=0$, which finally 
proves ellipticity of $L$. \hfill 
$\endproof$

The results obtained so far suffice to deduce an  
implicit-function-theorem.  To state it precisely, we need to 
choose appropriate function spaces. It is natural to choose 
Sobolev spaces since they are also used to show existence 
for the time evolution~[CY]. To begin with, it is convenient 
to summarize the order of spatial differentiation by which the 
various fields enter the quantities $\Gamma$, $U$, $\alpha$ and 
hence the GRTSE, by the following matrix:
$$
\def\linie{\vrule height 12pt depth 5pt width 0.4pt}
\def\back{\noalign{\vskip -2pt}}
\lineskip=0pt
\matrix{&\linie &  \Gamma & U & \alpha & \hbox{GRTSE}\cr
\noalign{\hrule}
g_{ab}               &\linie & 1 & 2 & 2 & 3 \cr\back
A_a^{\mu}            &\linie & 1 & 1 & 1 & 2 \cr\back
\phi^{\alpha}        &\linie & 1 & 1 & 1 & 2 \cr\back
{\dot g}_{ab}        &\linie & 0 & - & 0 & 1 \cr\back
{\dot A}_a^{\mu}     &\linie & 0 & - & 0 & 1 \cr\back
{\dot \phi}^{\alpha} &\linie & 0 & - & 0 & 1 \cr\back
\beta^a              &\linie & 1 & - & 1 & 2 \cr\back
\lambda^{\mu}        &\linie & 1 & - & 1 & 2 \cr}
\eqno{(3.15)}
$$
Note that we assumed that $U$ only contained first derivatives of the 
matter-fields, whereas it contains second derivatives of the 
gravitational field through the Ricci-scalar. Hence $g_{ab}$ enters   
the GRTSE thrice differentiated.\note{This seems 
to have been overlooked in [BF].}

By $H^n(V)$ we shall denote the Sobolev space of $V$-valued functions 
on $\Sigma$ with $L^2$-norm in the first $n$ derivatives 
(i.e. generalizing $H^n=W^{n,2}$ using an inner product in $V$). 
We shall have $V=T^0_2$ for $g_{ab}$ and ${\dot g}_{ab}$,
$V=T^0_1\otimes \LG$ for $A_a^{\mu}$ and ${\dot A}_a^{\mu}$,
$V=\reals^M$ for $\phi^{\alpha}$ and ${\dot \phi}^{\alpha}$,
$V=T^1_0$ for $\beta^a$ and $V=\LG$ for $\lambda^{\mu}$. The inner 
products for the various $V$'s are just as in the metric $G$, except 
for the gravitational field where instead of $G^{ab\, cd}$,
which is not positive definite, we choose the positive definite 
form $g^{ac}g^{bd}$ (compare (2.11)). Now we define the Sobolev
spaces
$$\eqalignno{
H^n_{\Phi}:=
&H^{n+3}(T^0_2)\times H^{n+2}(T^0_1\otimes \LG) 
               \times H^{n+2}(\reals^M),
&(3.16)\cr
H^n_{\dot\Phi}:=
&H^{n+1}(T^0_2)\times H^{n+1}(T^0_1\otimes \LG)
               \times H^{n+1}(\reals^M),
&(3.17)\cr
H^n_{\Psi}:=&H^n_{\Phi}\times H^n_{\dot\Phi},
&(3.18)\cr
H^n_{\xi}:=&H^n(T^1_0)\times H^n(\LG).
&(3.19)\cr}
$$
One may now show that the operator $F$ in the GRTSE, 
$F[\Psi,\xi]=0$, defines a $C^1$-map 
$$
F:\, H^n_{\Psi}\times H^{n+2}_{\xi}\rightarrow H^n_{\xi}
\quad\hbox{for}\, n\geq 2
\eqno{(3.20)}
$$
on the domain of fields $(\Psi,\xi)$ which satisfy Condition~1.\note{
The Sobolev embedding theorem for 3-dimensional domains and $L^2$-norms 
implies a continuous embedding $H^n(V)\hookrightarrow C^k(V)$ 
for $k< n-3/2$. $n\geq 2$ is needed to guarantee continuity of the 
functions and gain pointwise control, which is needed in the proof 
for $F$ being $C^1$.} For this we need to impose suitable but very
mild regularity conditions on the unspecified function $W$ in (2.5). 
The linear map $L$ is the first derivative of $F$ in wrt. the 
second argument: $D_2F[\Psi,\xi]$. Ellipticity implies that  
$\hbox{Image}(L):=L(H^{n+2}_{\xi})\subseteq H^n_{\xi}$ is closed and 
hence $H^n_{\xi}$ splits as orthogonal sum of closed subspaces, given 
by $L$'s image and the kernel of $L$'s adjoint. Hence, since $L$ 
is self-adjoint, $H^n_{\xi}=\hbox{image}(L)\oplus\hbox{kernel}(L)$. 
We now get an implicit function theorem for the map $F$ if 
$D_2F[\Psi,\xi]$ is a linear isomorphism, i.e. if 
$\hbox{kernel}(L)=\{0\}$. But since $L$ is elliptic, any non-trivial 
element in the kernel may be represented by a $C^{\infty}$ function 
$\xi'$ which must then satisfy (3.6). 
Hence a trivial kernel is equivalent to the following condition 
for smooth functions:

\noindent
{\bf Condition 2.}
$$
f_{\xi'}=\kappa\Gamma\ \hbox{implies}\  \xi=0,\ \kappa=0.
\eqno{(3.21)}
$$
Hence we arrive at the following formulation of an 
implicit-function-theorem for the generalized thin-sandwich problem.
It may be seen as the analog or generalization of Theorem~2 (and 3) 
in~[BF].

\proclaim Theorem. Let $n\geq 2$ and 
$(\Psi,\xi)\in H^n_{\Psi}\times H^{n+2}_{\xi}$ be a solution to 
the GRTSE, $F[\Psi,\xi]=0$, which satisfies Condition~1 
(i.e. (2.20-21)) and Condition~2 (i.e. (3.21)). 
Then there exist open neighbourhoods $V\subset H^n_{\Psi}$ of 
$\Psi$ and  $W\subset H^n_{\Psi}\times H^{n+2}_{\xi}$ of 
$(\Psi,\xi)$ and a $C^1$-map 
$\sigma:V\rightarrow H^{n+2}_{\xi}$ such that $F[\Psi',\xi']=0$ 
for $(\Psi',\xi')\in W \Leftrightarrow \Psi'\in V$ and 
$\xi'=\sigma(\Psi')$. \hfill

Consider the action $T$ of $\dot\reals:=\reals-\{0\}$ on 
$H^n_{\Psi}\times H^{n+2}_{\xi}$, given by 
$T_{\delta}(\Phi,\dot{\Phi},\xi):=(\Phi,\delta\dot{\Phi},\delta\xi)$ for 
$\delta\in\dot\reals$. It leaves individually invariant the three subsets 
of points $(\Psi,\xi)$ which 1.)~obey Condition~1, 2.)~obey Condition~2,
3.)~solve the GRTSE. To see this, recall that $f_{\xi}$ is linear in $\xi$,
hence $T_{\delta}\Gamma=\delta\Gamma$. Invariance of the first set 
is now obvious. Further, if $f_{\xi'}=\kappa\Gamma$ has only the trivial
solution, then so does $f_{\xi'}=\kappa T_{\delta}\Gamma$, since 
otherwise $(\delta^{-1}\xi',\kappa)$ would be a non-trivial solution 
to the first equation. Hence the second set is invariant. Finally,
since $\Gamma$ scales with $\delta$ and the square-root of expression 
(2.19) for $\alpha$ with $\vert\delta\vert$, 
the GRTSE (2.16) changes at most by an overall sign, which proves 
invariance of the third set.

We can now repeat the Theorem for each point $T_{\delta}(\Psi,\xi)$
on the $\dot\reals$-orbit of $(\Psi,\xi)$ with open sets $V_{\delta}$,
$W_{\delta}$ and solution maps $\sigma_{\delta}$. In this way the 
solution map $\sigma$ extends to a solution map 
$\sigma^*:V^*\rightarrow H^{n+2}_{\xi}$, where 
$V^*:=\bigcup_{\delta\in\dot\reals}V_{\delta}$, which uniquely 
represents all solutions in 
$W^*:=\bigcup_{\delta\in\dot\reals}W_{\delta}$. By construction it
satisfies 
$\sigma^*(\Phi,\delta\dot\Phi)=\delta\sigma(\Phi,\dot\Phi)$
$\forall\delta\in\dot\reals$. Dropping the superscript $*$, we
formulate this as 

\proclaim Corollary 1. Let $(\Psi,\xi)$ be as in the Theorem.
Then there exist open neighbourhoods $V\subset H^n_{\Psi}$ of
$\bigcup_{\delta\in\dot\reals} T_{\delta}\Psi$ and 
$W\subset H^n_{\Psi}\times H^{n+2}_{\xi}$ of 
$\bigcup_{\delta\in\dot\reals}T_{\delta}(\Psi,\xi)$ and a $C^1$-map
$\sigma:V\rightarrow H^{n+2}_{\xi}$ such that $F[\Psi',\xi']=0$ for 
$(\Psi',\xi')\in W \Leftrightarrow \Psi'\in V$ and $\xi'=\sigma(\Psi')$.
Moreover, 
$$
\sigma(\Phi,\delta\dot\Phi)=\delta\sigma(\Phi,\dot\Phi),
\quad\forall\delta\in\dot\reals
\eqno{(3.22)}
$$

Finally we prove that Condition~2 not only ensures local but also global
uniqueness. The analogous result has been proven for the traditional 
RTSE in [BO].\note{The full statement and proof given in 
[BO] contains an additional part which is erroneous, as was first 
pointed out in [BF]. If transcribed to our setting, the incorrect 
part would amount to the claim that (3.23) implied $r\equiv 1$.}

\proclaim Proposition 2. 
Let $(\Psi,\xi)$ and $(\Psi,\t{\xi})$ satisfy Condition~1
and $(\Psi,\xi)$ the GRTSE. Then $(\Psi,\t{\xi})$ satisfies 
the GRTSE $\Leftrightarrow$ there exists a positive function 
$r:\Sigma\rightarrow\reals_+$ such that $\Gamma=\dot\Phi+f_{\xi}$ 
and $\t{\Gamma}=\dot\Phi+f_{\t{\xi}}$ are related by 
$$
\t{\Gamma}=r\Gamma .
\eqno{(3.23)}
$$

\beginproof
$\Leftarrow$: This follows trivially from the fact that 
the GRTSE, i.e. equations (2.17-18), contain $\xi$ only 
through the combination $\overalpha\Gamma$.

\noindent
$\Rightarrow$: 
 For $s\in [0,1]$, consider the convex combinations 
$\xi(s):=s\xi+(1-s)\t{\xi}$ and $\Gamma_s:=\dot\Phi+f_{\xi(s)}=
s\Gamma+(1-s)\t{\Gamma}$. In the following it is useful to think of 
each $\Gamma_s$ as section in the pulled-back bundle $\Phi^*T(\Theta)$
whose fibre at $p\in\Sigma$ is a Minkowski space $\reals^{1,5+3N+M}$ 
with metric $G_{\Phi(p)}$. Condition~1 requires $\Gamma(p)$ and 
$\t{\Gamma}(p)$ to be ``timelike'', so that $s\mapsto\Gamma_s(p)$ is the 
straight path connecting these two ``timelike'' vectors. First we show 
that $\Gamma(p)$ and $\t{\Gamma}(p)$ lie in the interior of the same 
``light-cone'' for some, and hence all, $p\in\Sigma$. To see this,
we consider, for each $p$, the inner product $G(\Gamma-\t{\Gamma},{\cal V})$ 
with the timelike vector ${\cal V}:=g_{ab}{\partial\over\partial g_{ab}}$
of constant length-squared $G({\cal V},{\cal V})=-3$. Now,
$$
 \int_{\Sigma}G(\Gamma-\t{\Gamma},{\cal V})\,d\mu
=2\int_{\Sigma}\nabla_a(\beta^a-{\t{\beta}}^a)\,d\mu=0,
\eqno{(3.24)}
$$
so that, because $\Sigma$ is connected, there exists a point 
$p\in\Sigma$ where $G(\Gamma-\t{\Gamma},{\cal V})(p)=0$. 
Hence $\Gamma(p)$ and $\t{\Gamma}(p)$ point in the same half of the 
``light-cone'', and so does $\Gamma_s(p)$, since the interior of the 
half ``light-cone'' is a convex set. By continuity this must then be 
true at each point $p\in\Sigma$ so that $G(\Gamma_s,\Gamma_s)$ is a 
negative-valued function on $\Sigma$ for each $s$. 

Next we consider the function
$$
I(s):=S[\Psi,\xi(s)]=
\int_{\Sigma}\sqrt{-2U\,G(\Gamma_s,\Gamma_s)}\, d\mu.
\eqno{(3.25)}
$$
We have $I'(0)=0=I'(1)$, where $'={d\over ds}$, since $\xi$ and 
$\t{\xi}$ solve the GRTSE. Furthermore, a straightforward 
calculation yields:
$$
I''(s)=\int_{\Sigma}{[2U]^{1\over 2}\over 
[-G(\Gamma_s,\Gamma_s)]^{3\over 2}}\, 
G(\Gamma,\Gamma)G(\t{\Gamma}_{\perp},\t{\Gamma}_{\perp})\,d\mu 
\leq 0,
\eqno{(3.26)}
$$
with
$$
\t{\Gamma}_{\perp}:=\t{\Gamma}-
\Gamma{G(\Gamma,\t{\Gamma})\over G(\Gamma,\Gamma)}.
\eqno{(3.27)}
$$
The inequality in (3.26) results from $\Gamma$ being 
``timelike'' and $\t{\Gamma}_{\perp}$ being ``spacelike'' or zero.
But $I'(0)=I'(1)=0$ and $I''(s)\leq 0$ imply $I''\equiv0$.
On the other hand, equality in (3.26) can only be achieved for 
$\t{\Gamma}_{\perp}=0$ which is equivalent to (3.23), where $r$ 
must be positive-valued since $\Gamma$ and $\t{\Gamma}$ point in the 
same half of the ``light-cone''.\hfill
$\endproof$

Now (3.23) implies (3.6) with $\xi'=\t{\xi}-\xi$ and 
$\kappa={r-1\over r}$, so that Condition~2 will enforce 
$r=1$ and $\xi=\t{\xi}$. Hence we have

\proclaim Corollary 2.
If $(\Psi,\xi)$ and $(\Psi,\t{\xi})$ satisfy the GRTSE and Conditions~1 
and 2, then $\xi=\t{\xi}$.

\beginsection{Discussion}

It is obvious that the strategy of the (generalized) thin-sandwich 
approach cannot work for all data. Obvious bad data are those 
for which the Hamiltonian constraint cannot be solved for a nowhere 
vanishing $\alpha$. For example, consider fields $\Phi$ such that $U>0$ 
and velocities $\dot\Phi$ whose gravitational part is pure gauge:
${\dot g}_{ab}=2\nabla_{(a}{\xi'}_{b)}$. The Hamiltonian constraint 
implies $(\nabla_a\eta^a)^2\geq U\alpha^2$, where  $\eta=\xi'-\xi$, 
showing that $\alpha$ must vanish somewhere since 
$\int_{\Sigma}d\mu\,\nabla_a\eta^a=0$ and $\Sigma$ is connected. 
To avoid such situations, Condition~1 or its reversed version,  
$U<0$ and $G(\Gamma,\Gamma)>0$ may be imposed.
However, if the second condition is chosen formula (3.14) 
together with the proof of Lemma~3 show that $L$ manifestly fails  
to be elliptic, thus leaving only Condition~1. 

The technical Condition~2  has an interpretation in terms of the 
``canonical data'' $(\Phi,V)$, where $\alpha V=\dot\Phi+f_{\xi}$
and where we assume $(\Phi,\dot\Phi,\xi)$ to satisfy Condition~1
in order to have $\alpha\not =0$. 
Namely, if $f_{\xi'}=\kappa\Gamma$ for some non-zero 
$(\xi',\kappa)$, then $f_{\xi'}=\alpha\kappa V$ says that 
the same canonical data admit a representation in terms of 
the new lapse function $\alpha_{\rm new}=\kappa\alpha$ (now possibly
with zeros), gauge functions $\xi_{\rm new}=\xi'$ and coordinate-velocities 
${\dot\Phi}_{\rm new}=0$. Conversely, if an $\alpha_{\rm new}$ exists 
such that $\alpha_{\rm new}V=f_{\xi_{\rm new}}$, then 
$f_{\xi'}=\kappa\Gamma$ with $\kappa=\alpha_{\rm new}/\alpha$ and 
$\xi'=\xi_{\rm new}$. Hence Condition~2 precisely excludes the 
existence of other representations of the same canonical data with 
vanishing coordinate-velocities $\dot\Phi$. We note that Condition~2 
may itself be implied by simple geometric conditions on $\Phi$. 
One such set of conditions is provided by the following 

\proclaim Proposition 3. Condition~2 is implied by Condition~1 
and the following conditions on $\Phi$: 
$$\eqalignno{
&(i)\quad \hbox{$Ric$}<0 \ (Ric=\,\hbox{Ricci-tensor of $g$})
&(4.1)\cr
&(ii)\quad \nabla_a\lambda^{\mu}=0\ \hbox{and}\  
\lambda^{\mu}\rho_{\mu\beta}^{\alpha}\phi^{\alpha}=0
\ \hbox{imply}\ \lambda^{\mu}=0      
&(4.2)\cr}
$$

\beginproof
Let $f_{\xi'}=\kappa\Gamma$, then $G(\Gamma,\Gamma)<0\Rightarrow
G(f_{\xi'},f_{\xi'})\leq 0\Rightarrow$
$$
\int_{\Sigma}d\mu\left[G^{ab\, cd} 4\nabla_{(a}{\beta'}_{b)}
\nabla_{(c}{\beta}'_{d)}\right]
=\int_{\Sigma}d\mu\, 2\left[\nabla_{[a}{\beta'}_{b]}
\nabla^{[a}{\beta'}^{b]}-R_{ab}{\beta'}^a{\beta'}^b\right]\leq 0
\eqno{(4.3)}
$$
$\Rightarrow \beta'=0$. Since $G$ is positive definite on $f_{\xi'}$'s 
for which $\beta'=0$, $G(f_{\xi'},f_{\xi'})\leq 0$ implies $f_{\xi'}=0$
which for $\beta'=0$ is equivalent to the first two equations in 
(4.2).\hfill 
$\endproof$

\noindent
We recall that metrics with $Ric <0$ exist on any 3-manifold 
$\Sigma$~[GY] (e.g. quite in contrast to $Ric>0$, which is well 
known to imply a finite fundamental group). (4.2) should be 
read as a mild genericity-condition for the matter fields. 
 For example, if we have a single $U(1)$ gauge field and a charged 
scalar field (here represented by a real doublet $(\phi^1,\phi^2)$) 
then condition (4.2) is satisfied if the scalar field is not 
identically zero.

Finally we comment on the functional $S[\Psi,\xi]$ defined in (2.23). 
Given that Condition~1 is satisfied, solutions to the GRTSE are 
stationary points with respect to variations in $\xi$ (Lemma~1). 
Lemma~2 asserts that these must be minima which are stable if  
Condition~2 is satisfied. We now assume Conditions~1 and 2 to hold
and evaluate $S[\Psi,\xi]$ at a solution 
$\xi=\sigma(\Psi)$. We get a $C^1$-function 
$S_*: H^n_{\Psi}\rightarrow\reals_+$, 
$$
S_*[\Phi,\dot\Phi]=
\int_{\Sigma}d\mu\,\sqrt{-2U\,G\left(\dot\Phi+f_{\sigma(\Phi,\dot\Phi)},
\dot\Phi+f_{\sigma(\Phi,\dot\Phi)}\right)}\,,
\eqno{(4.4)}                        
$$
which satisfies 
$$
S_*[\delta\Psi]=\vert\delta\vert S_*[\Psi]
\eqno{(4.5)}
$$
for all $\delta\in\dot\reals$. Standard consequences are 
$$\eqalignno{
& D_2S_*[\Phi,\delta\dot\Phi](\dot\Phi)
=\hbox{sign}(\delta)\,S_*[\Phi,\dot\Phi],
&(4.6)\cr
& D_2^2S_*[\Phi,\delta\dot\Phi](\dot\Phi,\dot\Phi)=0,
&(4.7)\cr}
$$
where in (4.7) we assumed $C^2$-smoothness of $\sigma$.
It is tempting to try and regard $S_*$ as a kind of metric on at 
least an open subset of the tangent bundle of the space of 
fields. For pure gravity it generalizes a previously considered 
expression which is valid only for constant lapse function~[G] and 
also gives rigorous meaning to a formal definition of a distance 
function given in [CF]. But presently it is unclear to us whether 
$S_*$ indeed defines an interesting geometric 
structure.\note{One may wonder whether it defined a Finsler metric. 
For this one would have to show that the bilinear form 
$D_2^2S_*^2[\Phi,\dot\Phi]$ is (weakly) non-degenerate. But this 
is not even the case in finite dimensions for functions of the form  
(4.4) (i.e. sum of square-roots rather than square-root of sum). 
Take e.g. the function $f(y_1,\dots,y_n)=\sum_i\sqrt{y_i^2}$. Then
$\partial_i\partial_jf=2\partial_if\partial_jf$, which is obviously 
just of rank one.}

\vskip1.0truecm

\noindent
{\bf Acknowledgements:} I thank Julian Barbour, Robert Bartnik,
Gyula Fodor and Karel Kucha\v r for suggestions and clarifying 
remarks. This work was financially supported by NSF grant 
PHY95-14240 and a grant from the Swiss National Science 
Foundation.

\noindent

\beginsection{References}

\item{[BF]}
Bartnik, R., Fodor, G.: On the restricted validity of the thin
sandwich conjecture. {\it Phys. Rev.~D} {\bf 48}, 3596-3599 (1993)

\item{[BO]}
Belasco, E.P., Ohanian, H.C.: Initial conditions in general
relativity: Lapse and Shift Formulation. {\it Jour. Math. Phys.} 
{\bf 10}, 1503-1507 (1969)


\item{[BSW]}
Baierlein, R.F., Sharp, D.H., Wheeler, J.A.: Three-dimensional geometry
as carrier of information about time. {\it Phys. Rev.} {\bf 126},
1864-1865 (1962)

\item{[CF]}
Christodoulou, D., Francaviglia, M.: The geometry of the 
thin-sandwich problem. In: {\it Isolated Gravitating Systems in 
General Relativity}, edited by J. Ehlers. North-Holland,
New York (1979)

\item{[CY]}
Choquet-Bruhat, Y., York, J.W.: The Cauchy Problem. 
In: {\it General Relativity and Gravitation}, Vol~1, 
edited by A.~Held. Plenum, New York (1980)

\item{[G]}
Giulini, D.: What is the geometry of superspace?
{\it Phys.~Rev.~D} {\bf 51}, 5630-5635 (1995)

\item{[GY]}
Gao, Z.L., Yau, S.-T.: The existence of Ricci negatively
curved metrics on three manifolds. 
{\it Inv. Mat.} {\bf 85}, 637-652 (1986)

\item{[I]}
Isenberg, J.: Steering the universe. {\it Found. Phys} {\bf 16}, 
651-665 (1986)

\item{[K]}
Kiefer, C.: Non-minimally coupled scalar fields and the initial
value problem in quantum gravity. {\it Phys. Lett.~B} {\bf 225},
227-232 (1989)

\item{[W]}
Wheeler, J.A.: Geometrodynamics and the issue of the final state.
In {\it Relativity, Groups and Topology}, edited by B.~DeWitt and 
C.~DeWitt. Gordon and Breach, New York (1964)

\end